\documentclass[prl,twocolumn,showpacs,superscriptaddress,preprintnumbers,amsmath,amssymb]{revtex4}
\usepackage[latin1]{inputenc}
\usepackage{bm}
\usepackage{multirow,amssymb,amsbsy,amsmath}
\usepackage{graphicx}
\usepackage{verbatim}
\makeatletter
\usepackage{pifont}
\def\bra#1{\mathinner{\langle{#1}|}}
\def\ket#1{\mathinner{|{#1}\rangle}}

\usepackage{color}

\makeatother

\begin{document}



\title{Nonlocal memory effects allow perfect teleportation with mixed states}

\author{Elsi-Mari Laine}
\email{emelai@utu.fi}
\affiliation{Turku Centre for Quantum Physics, Department of Physics and 
Astronomy, University of
Turku, FI-20014 Turun yliopisto, Finland}

\author{Heinz-Peter Breuer}
\affiliation{Physikalisches Institut, Universit\"at Freiburg,
Hermann-Herder-Strasse 3, D-79104 Freiburg, Germany}

\author{Jyrki Piilo}
\affiliation{Turku Centre for Quantum Physics, Department of Physics and 
Astronomy, University of
Turku, FI-20014 Turun yliopisto, Finland}

\date{\today}

\begin{abstract}

We show that perfect quantum teleportation can be achieved with mixed photon polarization states when nonlocal memory effects influence the dynamics of the quantum system. The protocol is carried out with a pair of photons, whose initial maximally entangled state is destroyed by local decoherence prior to teleportation. It is demonstrated that the presence of strong nonlocal memory effects, which arise from initial correlations between the environments of the photons, allow to restore perfect teleportation. We further analyze how the amount of initial correlations within the environment affects the fidelity of the protocol, and find that for a moderate  amount of correlations  the fidelity exceeds the one of the previously known optimal teleportation protocol without memory effects. Our results show that memory effects can be exploited in harnessing noisy quantum systems for quantum communication and that non-Markovianity is a resource for quantum information tasks.

 \end{abstract}


\maketitle

One of the most striking consequences of quantum physics is quantum teleportation -- the possibility to transfer quantum states over arbitrary distances -- first proposed by Bennett {\it {et al}}.~in 1993 \cite{Bennett}. Since its theoretical introduction, teleportation has been demonstrated experimentally \cite{Bouwmeester, Boschi, Kim, Marcikic, Riebe, Barrett} up to the distance of 143 km \cite{Ma}. In the original proposal, the two parties, Alice and Bob, share a maximally entangled quantum state acting as a resource for the teleportation task. If, however, the maximally entangled state is influenced by noise and decoherence, perfect teleportation can no longer be accomplished \cite{Bennett, Verstraete}. Therefore, one of the current major challenges in accomplishing teleportation over long distances is to overcome the limitations imposed by decoherence and the subsequent mixedness of the resource state. 

When an open quantum system, due to its coupling with the external environment, 
continuously loses information to its surroundings, the noise induced dynamics is called 
Markovian  \cite{Lindblad, Gorini,Breuer2007}. Non-Markovian quantum dynamics with memory effects arise when the system does not only lose information, but temporarily recovers some of it from the environment at a later time \cite{BLP, BLP2}. Recently, a significant progress in developing a general theory of non-Markovian quantum dynamics \cite{BLP, BLP2,Piilo08,Vacchini08,Wolf, RHP, Kossakowski2010, Bassano1} as well as in the experimental detection and control of memory effects \cite{exp1,exp2,nmprobe} has been made.

In this Letter we show how quantum memory effects can be harnessed to give an advantage in mixed state quantum teleportation. We consider photonics realizations due to their dominant role in the experimental implementations of 
teleportation \cite{Ma}. The key element of the scheme introduced here are nonlocal memory effects where the local exposure of the bipartite quantum system to Markovian noise can create strong global memory effects \cite{nlnm}; a scheme which has also been recently experimentally demonstrated \cite{nmprobe}. The fundamental source for these effects are the initial correlations between the local environments of the bipartite open system. 
For entangled photon polarization states, nonlocal memory effects needed for the protocol arise naturally since the frequency distributions of the photons, which act as environments, are unavoidably correlated after a downconversion process \cite{nmprobe}. We also analyze how the amount of initial correlations between the environments affects the fidelity of the teleportation, and find that even in the absence of maximal correlations between the environments, the protocol will substantially outdo the performance of the known optimal protocol without memory effects \cite{Verstraete}.

In the standard quantum teleportation protocol Alice has a qubit whose state $\ket{\phi}_1=\alpha \ket{+}+\beta\ket{-}$ she wants to teleport to Bob. Alice and Bob share an entangled pair of qubits 2 and 3 in the Bell-state: 
\begin{equation} \label{eq1}
\ket{\phi^+}_{23}=\frac{1}{\sqrt{2}}(\ket{++}+\ket{--}).
\end{equation}
Alice performs a Bell-state measurement on the particles $1$ and $2$, which projects particle $3$ in Bob's hands into one of four states depending on which Bell-state Alice gains as an outcome. Alice further communicates her measurement outcome to Bob who performs a unitary operation on particle $3$ depending on the outcome of Alice.  The final state of particle $3$ is the original state $\ket{\phi}_3=\alpha \ket{+}+\beta\ket{-}$. During the Bell-state measurement particle $1$ becomes entangled with particle 2 and the state $\ket{\phi}_1$ is destroyed on Alice's side during the protocol.

\begin{figure}[tb]
\centering
\includegraphics[width=0.35\textwidth]{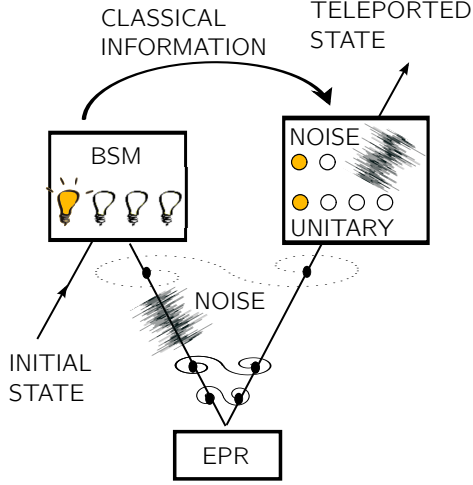}
\caption{\label{Fig:1} (color online). A schematic picture of the teleportation protocol with nonlocal memory effects. EPR refers to a source producing states of the form of Eq.~(\ref{eq1}) and BSM refers to a Bell-state measurement. The local noise before the BSM makes the EPR pair mixed while the noise in the last step of the protocol allows to recover the teleported state. }
\end{figure}

However, if the entangled pair of particles that Alice and Bob share is disturbed by noise, the fidelity of the standard teleportation goes down radically. Even if one chooses, instead of the standard teleportation scheme, an optimized protocol, perfect teleportation can no longer be achieved \cite{Verstraete}. Let us now demonstrate how perfect teleportation can be achieved with photon polarization states, even in the presence of noise, if the pair shared by Alice and Bob is influenced by nonlocal memory effects, see Fig.~\ref{Fig:1}.

Let us assume that the pair of entangled photons is created in a spontaneous parametric downconversion process after which the two particles are sent to remote locations to Alice and Bob. However, the particle sent to Alice is not perfectly isolated, but is interacting with its local environment giving rise to local decoherence, which destroys the entanglement  between the two parties. As a physical implementation of the local noise, we consider photon traveling through a quartz plate, where the polarization degree of freedom (system) and the frequency degree of freedom (environment)
interact \cite{exp2,nlnm,nmprobe}.

Now, Alice wishes to teleport the state $\ket{\phi}_1=\alpha \ket{H}+\beta \ket{V}$ to Bob. Here, $H$ ($V$) refers to horizontal (vertical) polarization state of the photon. Initially Alice and Bob share the Bell-state $\ket{\phi^+}_{23}=\frac{1}{\sqrt{2}}( \ket{HH}+\ket{VV})$ and the total initial state (system and environment) of the photons $2$ and $3$ is
\begin{equation}\label{eq2}
\ket{\psi(0)}=\ket{\phi^+}_{23}\otimes \int{d\omega_2 d\omega_3 g(\omega_2,\omega_3)\ket{\omega_2}\ket{\omega_3}}, \nonumber
\end{equation}
where $ g(\omega_2,\omega_3)$ gives the joint frequency amplitude of the photons $2$ and $3$ and $\int d\omega_2 d\omega_3|g(\omega_2,\omega_3)|^2=1$. 
The Hamiltonian for the local dephasing due to the quartz plate takes
the form
\begin{equation} \label{eq3}
 H_{i} = -\int d\omega_{i} \, \omega_{i} \Big[ n^i_V |V\rangle\langle V| +
 n^i_H |H\rangle\langle H| \Big]
 \otimes |\omega_{i}\rangle\langle\omega_{i}|, \nonumber
\end{equation}
where, e.g., $|V\rangle \otimes |\omega_{i}\rangle$ denotes the state of photon 
$i$ ($i=2,3$) with polarization $V$ and frequency $\omega_{i}$, and
$n^i_H$ ($n^i_V$) is the index of refraction for polarization component $H$ ($V$). 
The time dependent interaction Hamiltonian describing the evolution of the two-photon state is given by
\begin{equation}
H_I(t) = \chi_2(t) H_2+ \chi_3(t) H_3, \nonumber
\end{equation}
where
\begin{equation}\label{eq4}
 \chi_{i}(t) = \left\{ \begin{array}{ll}
 1 & {\mbox{if}} \;\;\; t_{\text{in}}^i \leq t \leq t_{\text{out}}^i \\
 0 & {\mbox{otherwise}}
 \end{array}\right. \nonumber
\end{equation}
and $t_{\text{in}}^i $ denotes the time photon $i$ enters the quartz plate and $t_{\text{out}}^i$  the time photon $i$ exits the quartz plate.  Now the time evolution of the total system is given by $ |\psi(t)\rangle = \exp\left[-i\int_0^t dt' H_I(t') \right] |\psi(0)\rangle$. Let us also write $t_i(t)=\int_0^t{\chi_{i} (t')dt'}$.  For convenience, in the following we will will not explicitly write the time dependence of $t_{i}$. 

\begin{figure}[tb]
\includegraphics[width=0.4\textwidth]{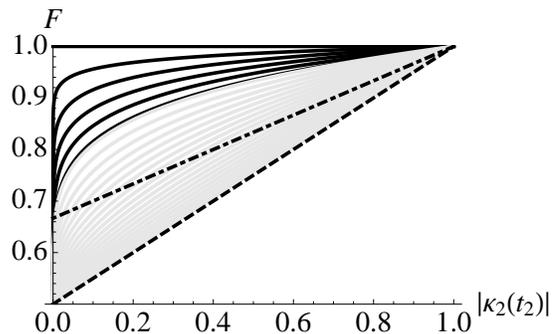}
\caption{\label{Fig:2} The fidelities of different teleportation protocols for the mixed state of Eq.~(\ref{eq5}) as a function of the decoherence function $|\kappa_2|$. The black dashed line represents the fidelity of the standard protocol $F^{\rm M}_{\rm w}$ [Eq.~(\ref{eq17})], the black dot dashed line the fidelity of the optimal protocol $F^{opt}$ [Eq.~(\ref{eq18})],  the black solid lines the fidelity of the protocol with memory effects  $F^{\rm NM}_{\rm w}$ with $\delta K\in[0,0.1]$ [Eq.~(\ref{eq16})] and  the grey solid lines with $\delta K\in[0.1,0.5]$.}
\end{figure}

In terms of our protocol, first photon 2 interacts with its quartz plate followed by the Bell-state measurement on photons 1 and 2. Then Alice communicates her measurement outcome to Bob and the interaction of photon 3 with its local environment in Bob's side finishes the procedure. The combined polarization state of the systems $2$ and $3$, after photon 2 has interacted with its quartz plate on Alice's side, is
\begin{eqnarray}\label{eq5}
\rho_{23}(t_2)&=&\frac{1}{2}(\ket{HH}\bra{HH}+\kappa_2(t_2)\ket{HH}\bra{VV} \nonumber\\
&&+\kappa_2^*(t_2)\ket{VV}\bra{HH}+\ket{VV}\bra{VV}),
\end{eqnarray}
where the decoherence function $\kappa_2$ is
\begin{equation}\label{eq6}
\kappa_2(t_2)=\int d\omega_2d\omega_3 |g(\omega_2,\omega_3)|^2 e^{-i \Delta n_2 \omega_2 t_2}, \nonumber
\end{equation}
$\Delta n_2=n^2_V-n^2_H$ is the birefringence, and $t_2$ the interaction time of 
photon 2.
Now, if Alice and Bob were to perform the standard teleportation with the shared state $\rho_{23}(t_2)$, the fidelity of the teleportation would decrease linearly with respect to $\kappa_2$, see Fig.~\ref{Fig:2}. Let us now describe how they can outperform this fidelity with nonlocal memory effects.

After the local interaction of photon $2$ the total state for the three photons is
\begin{equation}\label{eq7}
\ket{\Psi(t_2)}=\frac{1}{\sqrt{2}}\ket{\phi}_1(\ket{HH}\ket{\psi_{HH}(t_2)}+\ket{VV}\ket{\psi_{VV}(t_2)}), \nonumber
\end{equation}
where
\begin{eqnarray}\label{eq8}
\ket{\psi_{HH}(t_2)}&=&\int d\omega_2 d \omega_3 g(\omega_2,\omega_3)e^{i n_H^2\omega_2t_2}\ket{\omega_2}\ket{\omega_3},\nonumber\\
\ket{\psi_{VV}(t_2)}&=&\int d\omega_2 d \omega_3 g(\omega_2,\omega_3)e^{i n_V^2\omega_2 t_2}\ket{\omega_2}\ket{\omega_3}.\nonumber
\end{eqnarray}
This can be written in the form 
\begin{eqnarray} \label{eq9}
\ket{\Psi(t_2)}&=&\frac{1}{2}\ket{\phi^+}_{12} (\alpha\ket{H}_3\ket{\psi_{HH}(t_2)}+\beta\ket{V}_3\ket{\psi_{VV}(t_2)})\nonumber\\
&+&\frac{1}{2}\ket{\phi^-}_{12}  (\alpha\ket{H}_3\ket{\psi_{HH}(t_2)}-\beta\ket{V}_3\ket{\psi_{VV}(t_2)})\nonumber\\
&+&\frac{1}{2}\ket{\psi^+}_{12} (\beta\ket{H}_3\ket{\psi_{HH}(t_2)}+\alpha\ket{V}_3\ket{\psi_{VV}(t_2)})\nonumber\\
&+&\frac{1}{2}\ket{\psi^-}_{12} (\alpha\ket{V}_3\ket{\psi_{VV}(t_2)}-\beta\ket{H}_3\ket{\psi_{HH}(t_2)}), \nonumber
\end{eqnarray}
where 
\begin{eqnarray*}
 && \ket{\phi^+}=\frac{1}{\sqrt{2}}( \ket{HH}+\ket{VV}), \;\;\;
 \ket{\phi^-}=\frac{1}{\sqrt{2}}( \ket{HH}-\ket{VV}), \\ 
 && \ket{\psi^+}=\frac{1}{\sqrt{2}}( \ket{HV}+\ket{VH}), \;\;\; 
 \ket{\psi^-}=\frac{1}{\sqrt{2}}( \ket{HV}-\ket{VH})
\end{eqnarray*}
are the Bell states. Alice then performs the Bell state measurement and communicates her results to Bob.  Bob further applies a unitary operation on his particle depending on Alice's measurement outcome. Thus, so far, Alice and Bob have performed the standard teleportation scheme with the mixed state of Eq.~(\ref{eq5}). Now, in order to improve the protocol, Bob needs to harness the nonlocal memory effects. He can use the information sent by Alice and subject his particle to conditional noise depending on Alice's measurement outcome. By adding the conditional noise to his system, Bob actually cancels out the effect of the noise which earlier acted on Alice's system. He chooses the following unitary operations and the birefringence of his quartz plate to produce noise according to Alice's outcomes:
\begin{equation} \label{eq10}
\begin{array}{ccc}
\ket{\phi^+}& \Rightarrow& \mathbb{I}, \quad \Delta n_3=\Delta n_2 \\
\ket{\phi^-}& \Rightarrow &\sigma_z, \quad \Delta n_3=\Delta n_2 \\
\ket{\psi^+}& \Rightarrow& \sigma_x, \quad \Delta n_3=-\Delta n_2 \\
\ket{\psi^-}& \Rightarrow &i \sigma_y, \quad \Delta n_3=-\Delta n_2.
\end{array}
\end{equation}

Let us assume that Alice's outcome of the Bell measurement is the state $\ket{\phi^+}_{12}$. Now the total state of Bob's photon $3$ is
\begin{eqnarray}\label{eq11}
&&|\alpha|^2 \ket{H}\bra{H}\otimes\rho_{HH}+\alpha\beta^* \ket{H}\bra{V}\otimes\rho_{HV} \nonumber \\
&&+\alpha^*\beta \ket{V}\bra{H}\otimes\rho_{VH}+|\beta|^2 \ket{V}\bra{V}\otimes\rho_{VV}, \nonumber
\end{eqnarray}
where
\begin{eqnarray*} \label{eq12}
\rho_{HH}&=&\int d \omega_3 d\omega'_3 d\omega_2 g(\omega_2,\omega_3)g^*(\omega_2,\omega'_3) \ket{\omega_3}\bra{\omega'_3}=\rho_{VV}, \\
\rho_{HV}&=&\int d \omega_3 d\omega'_3 \tilde{g}(\omega_3,\omega'_3) \ket{\omega_3}\bra{\omega'_3} = \rho^{\dagger}_{VH},
\end{eqnarray*}
and $\tilde{g}(\omega_3,\omega'_3)=\int d\omega_2 g(\omega_2,\omega_3)g^*(\omega_2,\omega'_3) e^{-i \Delta n_2 \omega_2 t_2}$.
When Bob subjects his photon to noise by putting his photon through a quartz plate with $\Delta n_3=\Delta n_2$ his final state can be written as
\begin{eqnarray} \label{eq13}
\rho_{\rm F} &=&|\alpha|^2 \ket{H}\bra{H}+\alpha\beta^*\kappa(t_2,t_3) \ket{H}\bra{V}\\
&&+\alpha^*\beta \kappa^*(t_2,t_3) \ket{V}\bra{H}+|\beta|^2 \ket{V}\bra{V}, \nonumber
\end{eqnarray}
where the decoherence function is 
\begin{equation}
\kappa(t_2,t_3)=\int d \omega_2 d \omega_3 |g(\omega_2,\omega_3)|^2 e^{-i \Delta n_2 (\omega_2 t_2+\omega_3 t_3)},\nonumber
\end{equation}
and $t_3$ is the interaction time in Bob's quartz plate. 

In the description of the downconversion process, the frequency distribution can be taken to be a joint Gaussian 
distribution \cite{nlnm,nmprobe}
\begin{equation} \label{eq14}
 |g(\omega_2,\omega_3)|^2 = \frac{1}{2\pi\sqrt{\textrm{det}C}}
 e^{-\frac{1}{2}
 \left(\vec{\omega}-\vec{\langle \omega \rangle }\right)^TC^{-1}
 \left(\vec{\omega}-\vec{\langle \omega \rangle }\right)},
\end{equation}
where $C=(C_{ij})$ is the covariance matrix with elements $C_{ij}=\langle\omega_i\omega_j
\rangle-\langle\omega_i\rangle\langle\omega_j\rangle$. We assume that both 
the means and the variances of $\omega_2$ and $\omega_3$ are equal, i.e., $
\langle \omega_2 \rangle = \langle \omega_3 \rangle =\omega_0/2$ and $C_{11} = 
C_{22} = \langle\omega^2_i\rangle-\langle\omega_i\rangle^2$. To quantify the 
frequency correlations we use the correlation coefficient $K=C_{12}/\sqrt{C_{11}
C_{22}}=C_{12}/C_{11}$ satisfying $|K|\leq1$. 

If Bob chooses $t_3=t_2$ and the frequency distribution is of the form of Eq.~(\ref{eq14}) with the  correlation coefficient $K=-1$, the polarization state of photon $3$ is given by Eq.~(\ref{eq13}) with $\kappa(t_2,t_3)=e^{i \omega_0 \Delta n_2 t_2}$, i.e., the magnitude of the decoherence function has returned to its original value equal to 1. If Bob now performs a phase gate with phase $-\omega_0\Delta n _2 t_2$, perfect teleportation has been completed. The other measurement outcomes of Alice give the same result if Bob applies the operations given in Eq.~(\ref{eq10}). It is important to notice that all the steps of this protocol can be applied locally. 

We have demonstrated how Alice and Bob can recover perfect teleportation in the presence of noise if Bob after the standard teleportation protocol subjects his particle to noise. The key element in the protocol are the nonlocal memory effects which arise when the local decoherence is present on Bob's side. 
The source of the nonlocal memory effects lies in the initial correlations between the local environments described by the correlation coefficient $K$, which in turn is determined by the width of the original pump pulse in the downconversion process. In the case of perfect anticorrelation $K=-1$ (delta peak pump), perfect teleportation is achieved and for an uncorrelated distribution $K=0$ no advantage with respect to the standard protocol can be gained. It is also important to note that no initial entanglement is necessary between the local environments, but classical correlations suffice \cite{nlnm,nmprobe}.

Thus, we have shown that when the initial joint frequency distribution has perfect anticorrelation $K=-1$, this leads to perfect teleportation with mixed polarization states. If, however, the initial pump pulse does not have a delta peak distribution, the fidelity of the protocol decreases. In the following we study how a finite width of the pump pulse will affect the fidelity of the teleportation protocol. Let us take the correlation coefficient to be $K=-1+\delta K$, i.e., $\delta K$ measures the deviation from the ideal case. The fidelity  $F= {_1}{\bra{\phi}\rho_3\ket{\phi}}{_1} $ between the original state $\alpha\ket{H}+\beta\ket{V}$  and the teleported state becomes
\begin{equation} \label{eq15}
F=1-2|\alpha|^2|\beta|^2(1-|\kappa_2(t_2)|^{2\delta K}).\nonumber
\end{equation}
Considering the worst case scenario, i.e. $|\alpha|^2=|\beta|^2=1/2$, we obtain
\begin{equation}\label{eq16}
F_{\rm w}^{\rm NM}=1-\frac{1}{2}(1-|\kappa_2(t_2)|^{2\delta K}).
\end{equation}
If the teleportation was performed with the decohered state in Eq. (\ref{eq3}), using the standard scheme without taking advantage of the memory effects, we would have the fidelity
\begin{equation}\label{eq17}
F_{\rm w}^{\rm M}=1 - \frac{1}{2}(1 - |\kappa_2(t_2)|).
\end{equation}
If instead of the standard teleportation scheme, one would use an optimal teleportation scheme for the decohered state at hand \cite{Verstraete}, one would get the fidelity 
\begin{equation} \label{eq18}
F^{opt}=\frac{ |\kappa_2(t_2)|+ 2}{3}
\end{equation}
which gives the value $2/3$ for a classical state. The fidelities of the different teleportation protocols are plotted in Fig.~\ref{Fig:2} as a function of the decoherence function $|\kappa_2(t_2)|$. We see that the fidelity of the protocol with memory effects for sufficiently large values of the correlation coefficient $|K|$ is still much larger than the fidelity of the standard protocol or the optimal protocol in the absence of memory effects. For sufficiently large correlations ($\delta K\leq 0.1$, i.e. $-1.0 \leqslant K \leqslant -0.9$) the fidelity of the protocol with memory effects exceeds the fidelity of the optimal protocol without memory effects 
all the way to an almost fully decohered state. For a smaller value of correlations ($0.1\leq\delta K\leq 0.5$) the fidelity still exceeds the one of the standard protocol. It is also important to note that the high value of the correlation coefficient $K=-0.9$ (or $\delta K = 0.1$) is experimentally realizable \cite{nmprobe}.

Summarizing, we have found that nonlocal memory effects can substantially increase the fidelity of mixed state quantum teleportation. In the protocol Alice and Bob act on their particles locally and the nonlocal memory effects occur due to initial correlations between the local environments of the photons. In order for Bob to harness the memory effects, he needs to subject his photon to local noise after the standard teleportation protocol. We have shown that one can perform perfect mixed state teleportation of photon polarization states if the environments of the two photons share maximal initial correlations. The protocol presented here demonstrates how to overcome noise in a quantum communication setup by exploiting memory effects and that non-Markovianity is a resource for quantum information tasks. Moreover, this theoretical proposal can be implemented with existing technologies in an optical setup \cite{CF}.

\acknowledgments This work was supported by Academy of Finland (259827, mobility from Finland), Jenny and Antti Wihuri Foundation, Magnus Ehrnrooth Foundation, Turku University Foundation, the Graduate School of 
Modern Optics and Photonics, and the German Academic Exchange Service (DAAD).

\end{document}